# Three-dimensional artificial neural network model of the dayside magnetopause

A. V. Dmitriev and A. V. Suvorova

Skobeltsyn Institute of Nuclear Physics, Moscow State University, Moscow, 119899, Russia

**Abstract.** Artificial Neural Networks (ANN) from package NeuroShell 2 are applied for modeling of dayside magnetopause. The model data set is based on the magnetopause crossings and on the corresponding hour-averaged measurements of solar wind plasma and interplanetary magnetic field. ANN model represents 3D shape and size of the dayside magnetopause in dependence on solar wind dynamic pressure, $B_z$ and $B_y$ components of interplanetary magnetic field. The model permits firstly to describe dynamics of the magnetic cusp and global asymmetry of the dayside magnetopause under wide range of the external conditions. Slightly change of the magnetopause size with change of $B_y$ absolute value is presented quantitatively in the form of analytical expression. ANN model shows three regimes of the magnetopause dynamics that controlled by $B_z$ component: pressure balance regime ($B_z>0$ nT), reconnection regime ($0>B_z>-10$ nT) and regime of reconnection saturation ($B_z<-10$ nT). Three different regimes of magnetopause dynamics at the subsolar point are described by modified pressure balance equation obtained from ANN model.

## Introduction

Many empirical models of the magnetopause (MP) have been developed last years. Some investigators [*Formisano et al.,* 1979; *Sibeck, et al.*, 1991; *Kuznetsov et al.*, 1992, 1994] studied the empirical 3D magnetopause shape and size using general forms of conic surfaces (ellipsoid and paraboloid). They represented the model of the magnetopause surface as a function of the solar wind (SW) dynamic pressure under various orientation of interplanetary magnetic field (IMF). Hence these models represented only qualitative dependencies on the IMF. 2D MP models [*Roelof and Sibeck*, 1993; *Petrinec and Russell*, 1993, 1996; *Shue et al.*, 1997; *Kuznetsov and Suvorova*, 1996, *Kuznetsov et al.*, 1998] studied MP position dependence on the solar wind pressure and IMF more detailed. These models use simple elliptical, parabolic or trigonometric functions to provide a good fit of the MP shape at distances as far as 20 $R_E$ (Earth's radii) in the tailward direction. Two or more separate functional forms were used in some models, then the solutions were attached to each other in a piecewise continuous manner [*Formisano et al.,* 1979; *Petrinec and Russell*, 1996; *Kuznetsov and Suvorova*, 1994; *Kuznetsov et al.*, 1998]. 2D models assume an axial symmetry of MP surface, which offers a reasonably good first approximation. But this assumption can lead to certain errors because realistic magnetopause surface is asymmetrical in north-south and in dawn-dusk directions and it has such features as "neutral points" (or polar cusps) [*Formisano et al.,* 1979; *Sibeck et al.*, 1991; *Kuznetsov et al.* 1992,1994; *Petrinec and Russell*, 1995; *Kuznetsov and Suvorova*, 1997,1998a]. In general all previous models are based on three input parameters: 2 coordinates and 1 external parameter (3D models) or 1 coordinate and 2 external parameters (2D models).

In previous statistical studies it was established that the main parameters controlling the dayside MP are the dynamic pressure $P$ of the solar wind (SW) plasma and the negative $B_z$ component of the interplanetary magnetic field. According to the classical Chapman-Ferraro theory the magnetopause position may be derived from the pressure balance between the SW dynamic pressure $P$ and the pressure of the geomagnetic field $B^2/2\mu_0$, where $\mu_0$ is permeability of vacuum. Therefore MP distance $R$, at least at the dayside, should depend on $P^{-1/6}$. The important problem of the quantitative description of $B_z$ influence on the MP position is quite successfully solved with empirical modeling only.

The model by *Roelof and Sibeck* [1993] (here and after RS model) is the first complete empirical model based on the large statistics of MP crossings (795 of crossings have available SW and IMF hourly averaged data) by 12 high-apogee satellites. *Shue et al.* [1997] used for their model another data set of MP crossings including several geosynchronous satellite crossings with 5-min averaged values of $P$ and IMF $B_z$. The model data set contains 553 crossings acquired near the equatorial region. The model developed by *Kuznetsov and Suvorova* [1998b] is based on the data set of the RS model, with added geosynchronous crossings from data set compiled by *Kuznetsov and Suvorova* [1997; http://alpha.npi.msu.su/~alla] and on the hourly averaged SW and IMF data (842 MP crossings). All models are in well agreement at quiet and moderately disturbed conditions ($P<8$ nPa, $|B_z|<5$ nT) but yield different results in strongly disturbed periods.

The common characteristics of these empirical MP models are following: 1) an axial symmetrical magnetopause surface (2D) with apriory defined shape; 2) only two external parameters $P$ and $B_z$; 3) pressure balance equation is used as initial dependence on the external parameters. These restrictions are due to the regression method applying to the fitting procedure in the models. Note, the initial data is distributed strongly non uniformly in the phase space of the model parameters therefore an application of the regression method leads to limitation of the parameter number and dynamical range of the models.

Last time an alternative computational algorithm called Artificial Neural Network (ANN) is widely used for nonlinear multi-parameter data analysis and modeling. For instance, ANNs are applied to make short time forecasts of physical quantities varied in time [e.g. *Gorney et al.*, 1993; *Wu and Lundstedt*, 1996; *Dolenko et al.*, 1996] or to develop multi-parameter models [*Zhou and Yang*, 1996; *Lundstedt*, 1992, 1997; *Dmitriev and Orlov*, 1997a,b; *Dmitriev and Suvorova*, 1997,1998]. The main advantage of ANN method is that it permits to model a complex nonlinear physical system without any aprior knowledge about physical laws operating in the system. Due to the absence of the physical assumptions in ANN models the results of such kind of modeling may become "non-physical" and therefore doubted. In this situation the best solution is to test ANN technique, i.e. to apply ANN algorithm to the well-studied physical problem and to compare ANN results with well-known experimental facts and physical models.

The dynamics of shape and size of the Earth's magnetopause have not been studied yet by means of ANN. We use ANN package 'Neuroshell 2' [1996] to develop a complex 3D model of the dayside magnetopause. At the first stage the empirical MP model with a large number of input parameters is created by means of General Regression Neural Network with genetic search of the best smoothing factor (GRNN) [*NeuroShell 2*, 1996; *Caudill*, 1993]. At the second stage the Group Method of Data Handling (GMDH) [*Dolenko et al.*, 1996] is used to represent the model solution in the form of analytical expression. We have to emphasize that the ANN model is developed without any aprior assumptions about the magnetopause shape and about kind of its dependence on the external parameters. We assume only that the magnetopause shape has a mirror symmetry relatively to the ecliptic plane because insufficient statistics at high latitude in south hemisphere resulted in misrepresented of the shape of the magnetopause. We test the ANN model by means of comparison with previous models and results (theoretical and empirical) and we try to find some unknown effects.

## Initial Data

In the study we use the data set of the dayside magnetopause crossings from [*Roelof and Sibeck*, 1993] and geosynchronous crossings from [*Kuznetsov and Suvorova*, 1997; http://alpha.npi.msu.su/~alla]. The statistical distribution of the magnetopause crossings in the space of control variables $P$ - $B_z^{(GSM)}$ for both data sets mentioned above was studied in [*Kuznetsov et al.*, 1998]. Comparison of this distribution with total two-dimensional distribution of hourly averaged $P$ and $B_z$ values observed during 1964-1993 showed that the dynamic ranges of $P$ and $B_z$ at the moment of MP crossings cover nearly the whole range of physically reasonable parameter values: the pressure $P$ varies from 0.3 nPa up to 50 nPa and $B_z$ varies from -28 nT to 20 nT.

Here we discuss an important part of the data set - geosynchronous satellite crossings. There are 197 MP crossings by geosynchronous satellites from 1967 to 1993. Time intervals when the satellite going through magnetosheath varied from one minute to three hours. The SW dynamic pressure varies from ~2 up to 50 nPa. Obviously in very short events of the MP crossings we may observe only local MP deformation which is not expanded on the entire dayside magnetopause. The magnetopause may also reach unpredictably much smaller distances than 6.6 $R_E$ under very high dynamic pressure. To exclude such unsuitable observations of the geosynchronous MP crossings we apply for them the selection criteria developed in *Kuznetsov and Suvorova* [1997,1998a]. The minimal MP crossing duration was restricted in [*Kuznetsov and Suvorova*, 1998a] by six minutes that is about response time which is required for pressure discontinues with velocity about 500 km/s to sweep over the dayside magnetopause. There are 84 events with duration more than 6 minutes and available SW and IMF data: 13 events under positive $B_z$ and 71 events under negative $B_z$.

Fig. 1 (reprinted from [*Kuznetsov and Suvorova*, 1998a]) illustrates an application of the selection criteria on $P$ and $B_z$ to these 84 events. Under positive $B_z$ the MP crossings are observed in 13 events when the SW dynamic pressure is in the range of 17-29 nPa (squares on Fig. 1) that is in accordance to the pressure balance for the MP distances near 6.6 $R_E$. Under negative $B_z$ (open and dark circles) the dynamic pressure is varied widely from 1.7 nPa to 50 nPa. Earlier *Rufenach et al.* [1989] indicated that when the geosynchronous crossings occur for negative $B_z$, the MP position is reached at much lower $P$. *Kuznetsov and Suvorova* [1998a] have found a certain dependence between minimal observed pressures and longitudes of the magnetopause crossings. The minimal pressures are observed in 28 events (dark circles) under $B_z<-6$ nT. The minimal pressure smoothly increases from noon meridian to dawn or to dusk with clear jump from 4-6 nPa in prenoon sector to 10-15 nPa in postnoon one. Corresponding empirical dependencies were obtained and presented in Fig.1. Following to *Kuznetsov and Suvorova* [1998a] we use in our study only 41 magnetopause geosynchronous crossings: 13 events under positive $B_z$ and 28 events under large negative $B_z$ with relatively low SW dynamic pressure.

Finally the model data set contains 516 crossings of the dayside MP along with SW plasma and IMF parameters from OMNI Web and some physical parameters. ANN data set contains 999 MP crossings (we assume the mirror symmetry of the initial data relatively to the ecliptic plane). We consider 3 coordinate parameters of the MP location in GSE system (latitude $\lambda$, longitude $\varphi$ and distance $R$) and 17 external hourly averaged parameters: SW velocity ($V$), density ($n$), temperature ($T$) with their variations ($\delta V$, $\delta n$, $\delta T$), IMF strength $B$ with components in GSE ($B_x^{(GSE)}$, $B_y^{(GSE)}$, $B_z^{(GSE)}$) and in GSM ($B_y^{(GSM)}$, $B_z^{(GSM)}$) coordinate systems with their variations ($\delta B$, $\delta B_x$, $\delta B_y$, $\delta B_z$) and geomagnetic $Dst$-index. Using SW and IMF data we have calculated ten hourly averaged physical parameters: interplanetary electric filed components ($E_y \sim V \cdot B_z$, $E_z \sim V \cdot B_y$) and different kinds of SW pressure: dynamic $P = m_p n V^2$ ($m_p$ - proton mass), thermal $P_t \sim nkT$ ($k$ - Boltzmann constant), magnetic $P_m \sim B^2/8\pi$ and total $P_s = P + P_t + P_m$ with their variations ($\delta P$, $\delta P_t$, $\delta P_m$, $\delta P_s$).

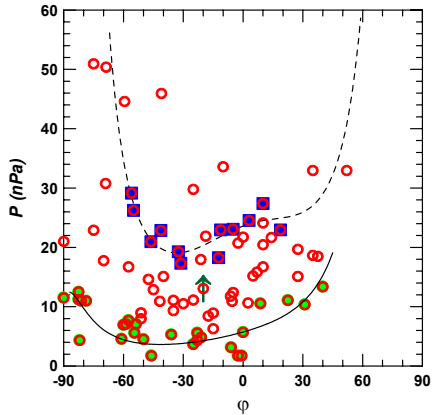

**Figure 1.** (reprinted from [*Kuznetsov and Suvorova*, 1998a]) Solar wind dynamic pressure versus MP crossing longitude during Bz>0 (solid squares), Bz<0 (open circles) and Bz<-6 (solid circles).

## ANN Simulation

*Normalization*

Data normalization procedure is necessary for successful operation of ANN. This procedure consists of two steps. The first one is investigation of probability distribution function of parameters to determinate the minimum, maximum and most probable parameter values. The second step is scaling of parameter values into the range from -1 to 1 with most probable value about zero. The analysis of the distribution functions of the SW and IMF parameters during 1964-1994 has shown [*Dmitriev and Orlov*, 1997b; *Veselovsky et al.*, 1998] that the parameters $B_x$, $B_y$, $B_z$ have a normal distributions and parameters $B$, $n$, $T$, $\delta B$, $\delta B_x$, $\delta B_y$, $\delta B_z$, $\delta T$, $\delta n$, $\delta V$ have a log-normal distributions. SW velocity $V$ has a complicated distribution function. The physical parameters of the solar wind (electric field components and pressures) including their hourly variations have log-normal distribution functions.

The distribution functions of parameters $B_z^{(GSM)}$, $B_y^{(GSM)}$, $n$, $V$, $P$ and $R$ are demonstrated in Fig. 2 (a-f). Dashed lines indicate the averaged distributions along the period 1964-1994; solid lines - the distributions for the model data set. As one can see in Fig. 2, the distribution functions for the model data set are similar to the averaged distributions. Therefore the SW and IMF data using in the model reflect adequately the real average conditions. We can see also in the Fig. 2 that $B_z$ and $B_y$ have the normal distributions. SW density and dynamic pressure have the log-normal distributions as mentioned above. The log-normal distribution of MP distance $R$ indicates a kind of preferable modeling function as logarithm of $R$ ($ln(R)$). All parameters with the log-normal distribution were recalculated in the logarithm scale.

*Model Parameterization*

The model parameterization is devoted to definition of the common type of modeling function and searching of optimal set of the variables, which permits to construct the model with sufficient accuracy. The functional dependence of $ln(R)$ on the other parameters is supposed to be a multiplication of functions.

To select the optimal set of the model parameters we use General Regression Neural Network (GRNN) with genetic search of best smoothing factor [*NeuroShell 2*, 1996; *Caudill*, 1993]. GRNN contains only one hidden layer. The number of neurons in the hidden layer is equal to number of examples in the training data set. This configuration of GRNN provides the best possible accuracy of the empirical model. The algorithm of genetic search of best smoothing factor permits to find the best solution of GRNN among the big population of partial solutions generated by big number of neurons in the hidden layer. Shortly GRNN is the best interpolator in the multidimensional space of the model parameters. The other important advantage of the algorithm is possibility to estimate the relative significance of each input variable numerically by means of the smoothing factor calculation for each parameter during ANN training mode. The training data set contains 60% (599 examples) of the initial data. Each of testing and examining data set contains 20% (200 examples) of the model data set. The output node of GRNN is $ln(R)$. At the beginning we use two angle coordinates, 28 parameters of SW and IMF and geomagnetic Dst-index as GRNN input nodes: $ln(R)=F\{\lambda, \varphi, ln(B), B_x^{(GSE)}, B_y^{(GSE)}, B_z^{(GSE)}, B_y^{(GSM)}, B_z^{(GSM)}, ln(T), ln(n), V, ln(\delta B), ln(\delta B_x), ln(\delta B_y), ln(\delta B_z), ln(\delta T), ln(\delta n), ln(\delta V), ln(E_y), ln(E_z), ln(P), ln(P_t), ln(P_m), ln(P_s), ln(\delta P), ln(\delta P_t), ln(\delta P_m), ln(\delta P_s), Dst\}$. In this case we obtain the best accuracy of the model with correlation coefficient $c \cong 0.99$. Then we repeat this procedure with those parameters that have the most relatively significance. After the completion of the parameter selection only five most significant input parameters were left: $\lambda, \varphi, B_y^{(GSM)}, B_z^{(GSM)}, ln(P)$, providing the accuracy of the model $c \cong 0.92$ and standard deviation 1.04 $R_E$.

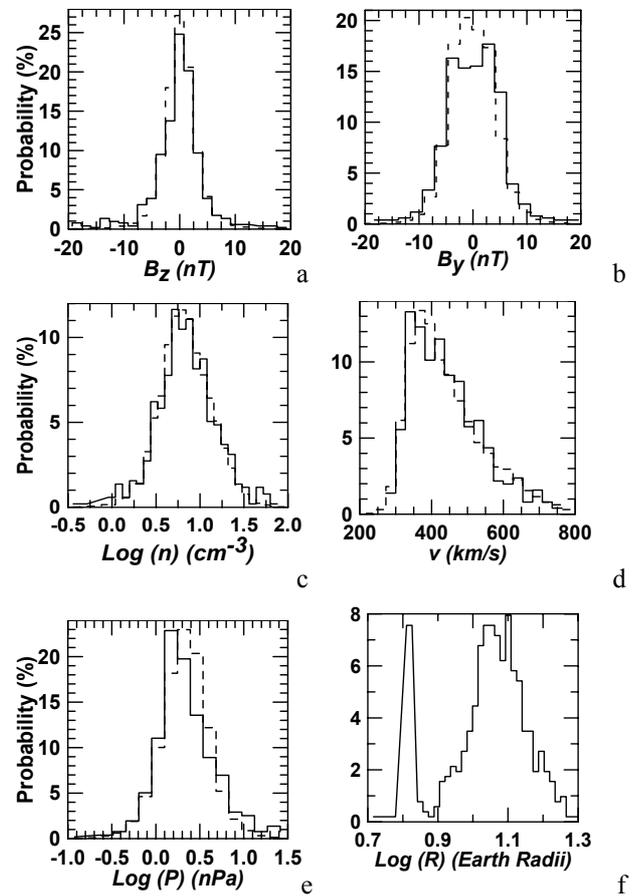

**Figure 2.** Statistical distributions of $B_z$ (a) and $B_y$ (b) components of IMF, solar wind density $n$ (c), velocity $V$ (d), dynamic pressure $P$ (e) and Earth-magnetopause distance $R$ (f) for the data along the period 1964-1994 (dashed lines) and for the model data set (solid line).

GRNN model generates a very complicated long program code, which does not be simplified. The model can be used only under the NeuroShell 2 core that is very difficult in routine calculations. To generate a portable simple program code of the model we apply ANN Group Method of Data Handling (GMDH) to the results of GRNN model calculations. GMDH contains one hidden layer with variable number of neurons. The number of neurons is defined by the number of polynom's terms that is searched during the training mode. The main advantage of GMDH is capability to represent the model code in the analytical form of polynom up to 3rd order. Unfortunately the accuracy of GMDH is moderate due to relatively low polynom order. This problem is solved by using simple functions together with the model parameters as input nodes of GMDH model.

*Partial Function Search*

To increase the accuracy of GMDH model we perform additional analysis in order to find the best simple functional representation of each input parameter of the model. The set of searching functions includes: *exp*, *ln* (natural logarithm), *pow* (power), *sin*, *cos*, and polynom of $k$-order ($P^k$, $k<5$). 5D matrix $M_R \equiv [R(\lambda_i, \varphi_j, B_{y_k}, B_{z_l}, ln(P)_n)]$ of magnetopause distances $R$ was calculated by means of GRNN model. The values of input nodes are varied in the following ranges: $\lambda=0 \div 80°$ (8° grid); $\varphi=-90 \div 90°$ (9° grid); $B_y=-20 \div 20$ nT (10 nT grid); $B_z=-20 \div 10$ nT (5 nT grid); $ln(P(nPa))=-0.3 \div 3.7$ (0.8 grid). The parameters values in the grid points ($i=1..11$, $j=1..21$, $k=1..5$, $l=1..7$, $n=1..6$) are used to calculate the matrix $M_R$ by means GRNN model. In result we generate the artificial data set which contains $11 \times 21 \times 5 \times 7 \times 6 = 48510$ examples of magnetopause location under any considering condition.

Linear regression of each functional representation for a single variable (the values of other variables are fixed) is used to find the best one chosen on the basis of the highest correlation between GRNN model dependence and considered function. This procedure is repeated for all number of different values of variables. As a result we find the types of functions from the considered set that are the most suitable to describe the model dependence on each parameter.

After the completion of Partial Function Search procedure the following additional GMDH input nodes were obtained: $\lambda^2$, $\lambda^3$, $sin(4\lambda)$, $cos(4\lambda)$, $sin(5\lambda)$, $cos(5\lambda)$, $\varphi^2$, $\varphi^3$, $sin(\varphi/2)$, $cos(\varphi/2)$, $sin(2\varphi)$, $cos(2\varphi)$, $sin(3\varphi)$, $cos(3\varphi)$, $B_y^2$, $B_y^3$, $ln(13+B_z)$ and $exp(ln(P))$. Thus we have selected five most significant parameters of MP model and we have searched 18 additional partial functions of these parameters as GMDH input nodes.

*Polynomial Solution of GMDH Model*

The input nodes of GMDH model are following: $\lambda$, $\lambda^2$, $\lambda^3$, $sin(4\lambda)$, $cos(4\lambda)$, $sin(5\lambda)$, $cos(5\lambda)$, $\varphi$, $\varphi^2$, $\varphi^3$, $sin\varphi/2$, $cos(\varphi/2)$, $sin(2\varphi)$, $cos(2\varphi)$, $sin(3\varphi)$, $cos(3\varphi)$, $B_y$, $B_y^2$, $B_y^3$, $B_z$, $ln(13+B_z)$, $P$, $ln(P)$. The matrix $M_R$ were used as GMDH output data set. The training data set contains 60% (29106 examples) of artificial data set. The testing and examining data sets divide equally the rest of 40% (9702+9702 examples). The ANN training time was about 10 hours on PentiumPro-200 (256kb cash, 64Mb RAM). The best solution of GMDH network (correlation coefficient $c \cong 0.83$) has long polynomial expression for magnetopause distance $R(\lambda, \varphi, B_y, B_z, P)$ presented on the www-page http://alpha.npi.msu.su/~alla

The comparison between the observed $R_{exp}$ and the modeled $R_{mod}$ magnetopause radial distances is presented in Fig. 3. The solid line represents exact agreement of $R_{mod}$ with $R_{exp}$. The accuracy of the model is defined as standard deviation $\sigma = \sqrt{\sum (R_{exp} - R_{mod})^2 / N}$ (where N=516 is total number of initial data points) and equal to $1.44 R_E$. Magnetopause crossings on the flank $|\varphi|>80°$ ($N$=56) are indicated by the crosses, crossings at very small dynamic pressure $P$<1 nPa ($N$=51) are indicated by circles, and other crossings are indicated by triangles. As we see in Fig.3 ANN model underestimates the magnetopause distances in the range of $R$>15$R_E$. Most magnetopause crossings at relatively large distances are observed on the flank. The number of crossings at very low pressure at distances $R>15R_E$ is significantly smaller. Therefore, ANN model predicts smaller distances than observed ones on the flanks of the magnetopause.

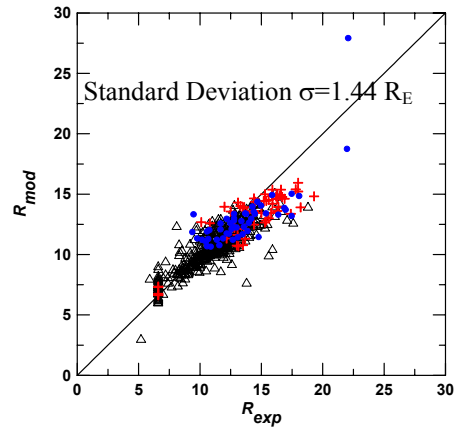

**Figure 3.** The comparison between the observed $R_{exp}$ and the modeled $R_{mod}$ magnetopause radial distances (in the Earth's radii). The solid line represents exact agreement of $R_{mod}$ with $R_{exp}$.

To compare quantitatively the ANN model accuracy in different regions of magnetopause and under different external conditions we use an average relative deviation δ of the model from the measurements: $\delta(\%) = \frac{100}{N} \sum_{k=1}^{N} \frac{|R_{exp} - R_{mod}|}{R_{exp}}$. Table 1 shows the average relative deviations in different MP angle sectors that are characterized by angle $\theta$ between X-axis and radius-vector to the magnetopause. The number of experimental measurements in the sector is presented in the middle column of the Table 1. ANN model accuracy is decreased from δ=6.6% in the noon sector ($\theta=0°\div10°$) to δ=11% at large angles $\theta=80°\div90°$. As discussed above the lager errors within angle sector $\theta=80°\div90°$ correspond to the flank area. To increase the accuracy of ANN model on flank region we need to expand the model toward the MP tail. However, the experimental data in this region are insufficient to develop 3D model with sufficient precise.

**Table 1.** ANN model accuracy in different angle sectors of the magnetopause

| $\theta$ (°) | N | $\delta$ (%) |
|---|---|---|
| 0÷10 | 21 | 6.61 |
| 10÷20 | 36 | 9.31 |
| 20÷30 | 68 | 7.48 |
| 30÷40 | 65 | 8.1 |
| 40÷50 | 46 | 8.3 |
| 50÷60 | 65 | 9.63 |
| 60÷70 | 70 | 9.57 |
| 70÷80 | 61 | 8.45 |
| 80÷90 | 84 | 10.9 |

**Table 2.** ANN model relative deviation δ (%) under different external conditions.

| | P (nPa) | | | | |
|---|---|---|---|---|---|
| $B_z$ (nT) | 0.3÷1 | 1÷2 | 2÷4 | 4÷8 | 8÷40 |
| 3.5÷20 | 10(6) | 12(16) | 5.6(14) | 9.2(12) | 8.2(9) |
| 0÷3.5 | 11(18) | 8.4(83) | 9.2(73) | 6.9(12) | 11(6) |
| -3.5÷0 | 8.2(28) | 8.1(70) | 8.3(57) | 8.9(16) | 14(5) |
| -7÷-3.5 | | 8.8(19) | 9.6(21) | 15(7) | 11(2) |
| -20÷-7 | 5.7(4) | 15(3) | 6.1(4) | 10(13) | 8.9(10) |

\* The numbers within round brackets are the numbers of experimental measurements

The ANN model accuracy dependence on the external conditions (P and $B_z$) is presented on Table 2. The numbers within round brackets are the numbers of experimental measurements in the given $P$-$B_z$ bin. In common the averaged relative deviation is better than δ=10%. For quite conditions $P<2$ nPa, $B_z>-3.5$ nT ANN accuracy is about δ~8% except cases of very low pressure ($P<1$ nPa) and large positive $B_z>3.5$ nT when the model accuracy is reduced to δ~11%. Under moderate SW conditions ($2<P<8$ nPa, $-7<B_z<-3.5$ nT) the averaged accuracy of the model is δ~9%. ANN model accuracy is about δ~10% for the "shock" events when the magnetosphere is disturbed by strong pressure pulse ($P>8$ nPa, $B_z>-7$ nT) and for the strong disturbed conditions associated with a large negative $B_z<-7$ nT.

*Comparison with previous models*

Comparison of the accuracy of the different MP models shows that the standard deviations are practically the same for all models: σ~1÷2 $R_E$ in RS model, σ=1.24 $R_E$ in the model by *Shue et al.* [1997], σ=1.55 $R_E$ in *Kuznetsov et al.* [1998] model and σ=1.44 $R_E$ in ANN model. Here we would like to remind that the model by *Shue et al.* [1997] was developed using data set restricted by low latitudes. Therefore actually this model may over estimate the MP distance at high latitudes because the dayside MP is flattened into an oblate ellipsoid [*Sibeck et al.*, 1991; *Kuznetsov et al.*, 1992].

We emphasis that the accuracy (σ≥1 $R_E$) is similar for the four models that use different data sets (both hourly and minute averaged) and developed by different methods. This fact may be associated with uncertainties in the initial data sets of the models that reflect features of the magnetopause dynamics. We suppose that these uncertainties are due to permanent oscillations of the magnetopause surface - "surface waves" due to the Kelvin Helmholtz instability [Dungey, 1955] and "standing waves" supposed in [*Kuznetsov and Suvorova*, 1998b]. Reported characteristics of the magnetopause waves propagating tailward [*Saunders*, 1989; *Chen et al.*, 1993] have wavelength of several $R_E$, periods about 5 min and amplitude of about 1 $R_E$. *Seon et al.*, [1995] shown that the magnetopause tends to be more oscillatory during the condition of solar wind velocity higher than ~400km/s. Analysis of the statistical distribution of SW velocity (Fig. 2d) shows that the conditions with $V>400$ km/s are about twice frequently than that with $V<400$km/s. Obviously the existence of the surface waves with described above characteristics may limit the accuracy of the empirical magnetopause models by 1 $R_E$.

## Discussion

*Cross sections of 3D dayside magnetopause*

The results of MP shape model calculations for the various SW and IMF conditions are presented in Fig. 4 (a-c), where left panels correspond to meridian cross-sections, and right panels correspond to equatorial cross-sections in the GSE coordinate system. One can see that MP size in equatorial plane is larger than in the meridian plane under the same external condition. This result agrees with previous studies [*Sibeck et al.*, 1991; *Kuznetsov et al.*, 1992]. The magnetopause distance $R$ decreases with increasing of SW dynamic pressure $P$ (Fig. 4a) and with decreasing of $B_z$ IMF component (Fig. 4b), that confirms the main earlier results.

On the meridian sections of the magnetopause (left panels in Fig. 4(a-c)) it is clearly seen "dimples" at latitudes $\lambda \approx \pm 45°$, that are associated with magnetic cusp regions. Similar features were represented in 3D model of *Formisano et al.* [1979] as the cusp region, which located at latitudes as higher than 64°. Note that in this study the cusp location is directly associated with fitting procedure: intersections between the dayside and tailside fitting surfaces to the magnetopause. Another "dimple" near equator plane $Z=0$ arises at $B_z<-10$ nT (left panel in Fig. 4b). We interpret this feature as magnetospheric magnetic field erosion in subsolar region under the strong negative $B_z$ [*Rufenach et al.*, 1989].

The equatorial section dynamics dependence on the SW and IMF parameters is presented on the right panels of Fig. 4. The strong deformations of the MP shape are clearly seen: the shape is asymmetrical relatively $X$-axis under any SW and IMF conditions. The magnetopause distance on the flank (along $Y$-axis) in the dawn sector ($R_{y-}$) is larger than in the dusk one ($R_{y+}$). We use this difference to estimate the asymmetry of the magnetopause as angle $\psi = \arccos\left(\sqrt{R_{y-}/R_{y+}}\right)$. If the MP is described by axial-symmetry curve (parabola or ellipse) then the angle $\psi$ will be approximately equal to deviation angle $\psi_a$ of MP axis of symmetry from $X$-axis. Table 3 shows the angles $\psi$ under different external conditions ($P$ and $B_z$). The angle is negative under any condition, i.e. MP axis is oriented in prenoon sector. The angle varies from $-14°\div-24°$ at quite conditions to $-28°\div-39°$ under disturbed conditions. We can indicate the tendency of asymmetry angle increasing with SW dynamic pressure. The asymmetry angle dependence on $B_z$ IMF component is non-regular. In average (under normal condition $P\sim1.5$ nPa, $B_z\sim0$ nT) the deviation angle is about $\psi=-24°$.

There were different opinions on the MP asymmetry in previous theoretical and empirical investigations. All these studies suggest that the asymmetry degree should depend at least on the interplanetary medium characteristics and Earth's motion around the sun. Some previous empirical MP studies [*Fairfield*, 1971; *Kuznetsov and Suvorova*, 1998b] did not show significant asymmetry (>1 $R_E$) of the dayside magnetopause near the equatorial plane at quiet and moderate conditions. On the other hand it was found small asymmetry ($\psi_a\sim-6.6°$) at moderate SW dynamic pressure $P\sim3$ nPa [*Formisano*, 1979]. Indirect indications on significant MP asymmetry ($\psi_a\sim-20°$) under disturbed SW conditions that lead to strong compression or/and erosion of the magnetospheric magnetic field were found in [*Rufenach et al.*, 1989; *Kuznetsov and Suvorova*, 1997; 1998a]. But results of [McComas *et al.*, 1994] argued against this conclusion. We have to remind that for the axial-symmetrical MP models the asymmetry is defined as angle $\psi_a$ of the deflection of MP axis of symmetry from $X$-axis. The theoretical studies showed different degree of the MP asymmetry. *Walters* [1964] predicts that the angle of the MP asymmetry varied from $\psi_a\sim-3°$ up to $\psi_a\sim-22°$ in dependence on IMF orientation and on ratio between the SW plasma pressure and IMF pressure. *Zhuang et al.* [1981] have found small MP asymmetry. The ANN modeling shows that magnetopause has large asymmetry under any external conditions. It seems there are a few reasons for the existence of the MP asymmetry, which may be associated both with SW and IMF characteristics and with properties of the Earth's magnetosphere. We will continue to discuss this question in the next section.

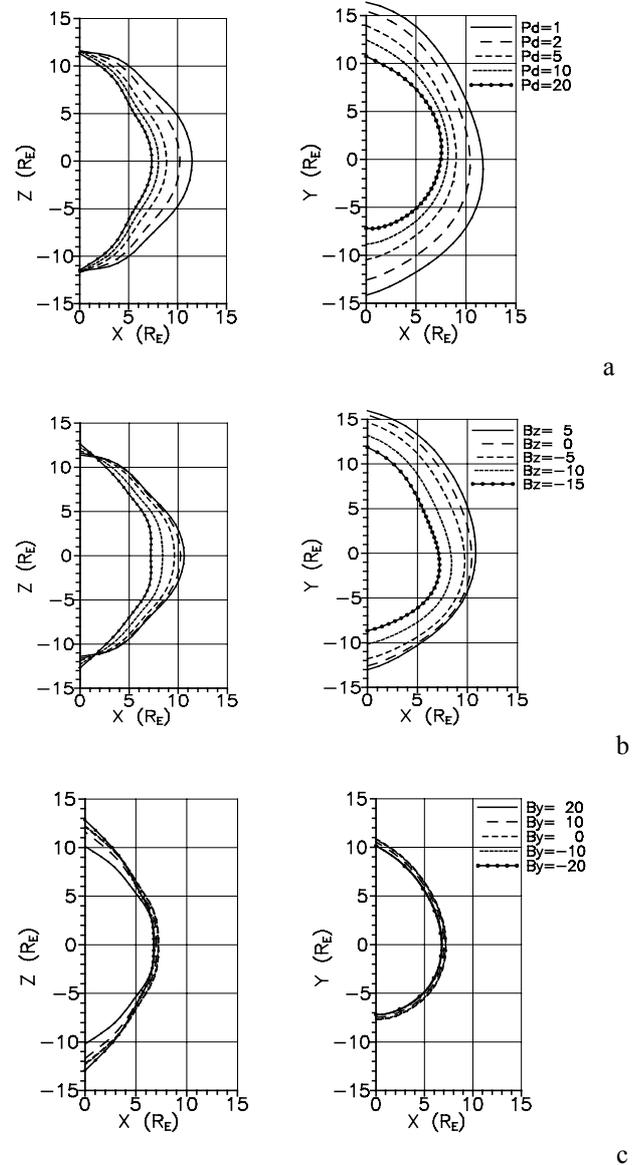

Fig. 4. Numerical ANN magnetopause surface model in solar-ecliptics coordinate system (GSE). Left panels are magnetopause sections in meridian plane ($XZ$); right panels are magnetopause sections in ecliptic plane ($XY$). The model is computed for a) $B_y=0$, $B_z=0$, $P=1, 2, 5, 10, 20$; b) $B_y=0$, $P=2$, $B_z=5, 0, -5, -10, -15$; c) $P=10$, $B_z=-10$, $B_y=20, 10, 0, -10, -20$.

**Table 3.** Magnetopause asymmetry angles $\psi$ under different external conditions

| $B_z$ (nT) | $P$ (nPa) | | | | |
|---|---|---|---|---|---|
| | 0.5 | 1.5 | 5. | 10 | 30 |
| 20 | -15° | -24° | -31° | -34° | -39° |
| 10 | -14° | -24° | -30° | -33° | -37° |
| 0 | -18° | -24° | -29° | -33° | -36° |
| -10 | -24° | -27° | -31° | -33° | -35° |
| -20 | -28° | -29° | -31° | -32° | -33° |

*IMF $B_y$ influence on the dayside magnetopause*

ANN model allows firstly estimate the role of $B_y$ IMF component in the MP dynamics. It is well known that direction of the $B_y$ strongly influence on such phenomena and processes at high latitude ionosphere as position of the magnetic cusp projection and electric current distribution [e.g. *Candidi et al.*, 1989; *Wilhjelm et al.*, 1978]. $B_y$ influence on the MP size and shape was still an open question. ANN model permits to solve this problem empirically. In GRNN model $B_y$ component has the third position in 'external' parameter significance after the $P$ and $B_z$. We present here an expression extracted from the whole formulae for $R$ in GMDH model and containing the explicit $B_y$-dependence:

$$R = F(\lambda,\varphi,B_z,P)/\exp\{0.033 + 5.9*10^{-6} f(\lambda,\varphi,B_y,B_z,P)\} \quad (1)$$

$$f(\lambda,\varphi,B_y,B_z,P) = B_y[273. - 45.1*(1-\varphi^2/4050)(1.55P^{1/4} - 2.37)] +$$
$$+ B_y^2[4.25 - 2.3*(1-\varphi^2/4050)(1.55P^{1/4} - 2.37)] +$$
$$+ B_y^2[1.83*(5.56\ln(P) - 7.72) - 6.78*(0.1B_z + 0.257)] +$$
$$+ B_y^3*(\lambda^2/311. - 7.5)/10. - B_y^4*0.0147$$

where $F(\varphi,\lambda,B_z,P)$ is a function of $\lambda$, $\varphi$, $B_z$, and $P$. Despite of complicated expression we can conclude that the influence of $B_y$ value on the magnetopause size is relatively small.

Fig. 4c demonstrates the behavior of MP in dependence on $B_y$ under disturbed SW and IMF conditions: large dynamic pressure $P$=10 nPa and strong negative $B_z$=-10 nT. As we can see in Fig. 4c the MP size decreases slightly with increasing $B_y$ absolute value (<0.5 $R_E$). The shape of the equatorial MP is practically not change with $B_y$. Indeed the equation (1) contains only square of the longitude ($\varphi^2$). Therefore we can conclude that $B_y$ does not affect on the asymmetry of the dayside MP.

This result is significant in the problem of the MP dawn-dusk asymmetry origin. Theoretical study of *Walters* [1964] proposed that the asymmetry is associated with external agent such as oblique IMF (the oblique angle between the spiral IMF and radial SW stream) with MP axis directed toward the dawn side. It means that the asymmetry is independent on $B_y$ orientation under undisturbed interplanetary conditions. Recently theoretical study of reconnection on MP [*Dreher and Schindler*, 1997] indicated intrinsic reason of the asymmetry of phenomena on the magnetopause and it showed independence of asymmetry on sign of $B_y$. The localized MP reconnection was simulated with Hall-MHD model and it was revealed that the Hall term in the Ohm's Law introduces a dawn-dusk asymmetry in MP phenomena. The authors showed that in the early phase of reconnection process magnetic flux is transported to the dawn-side irrespective of the sign of $B_y$ IMF component in the magnetosheath. They conclude that the early flux transport by the electron motion may account for dawn-dusk asymmetries in the occurrence of flux transfer events as well as in other magnetopause phenomena. Therefore, our result about independence of the MP asymmetry on $B_y$ is in accordance to the theoretical conclusions. We suggest that MP asymmetry at normal conditions ($B_z$>0 nT) is explained by external source (IMF) and the intrinsic source is responsible for the MP asymmetry under disturbed conditions ($B_z$<0 nT).

In Fig. 5 we present the subsolar point ($Y$=0, $Z$=0) distance $R_{ss}$ calculated by ANN model separately for $B_y$=0 (solid lines) and for $B_y$=20 nT (dashed lines) in $P$-$B_z$ coordinates. We clearly see the difference between the subsolar point locations calculated for $B_y$=0 nT and for $B_y$=20 nT. The difference decreases in the range of low dynamic pressure ($P$<1 nPa) and it increases at higher dynamic pressures. The difference can not be simply explained by contribution of the magnetic pressure into the total pressure value. At low dynamic pressure the effect is opposite to the expected one: the MP size increases with $B_y$.

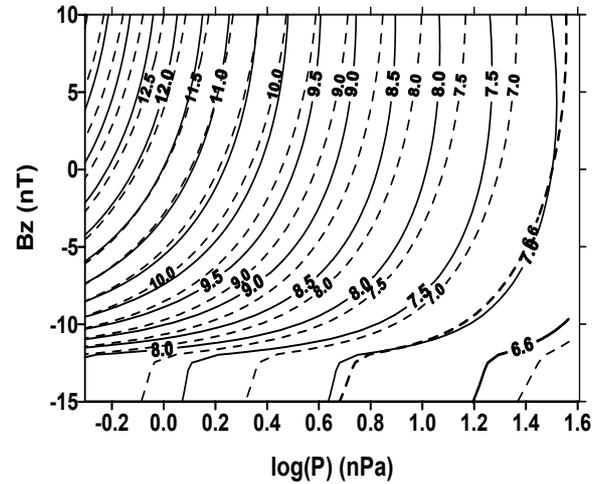

**Figure 5.** The subsolar point distance (in the Earth's radii) as a bivariate function of solar wind dynamic pressure $P$ and IMF $B_z$ component for $B_y$=0 nT (solid lines) and for $B_y$=20 nT (dashed lines).

To decrease the MP distance on 10% the total pressure must increase at last about twice. At high dynamic pressures ($P$>2 nPa) the IMF magnetic pressure value is much smaller than dynamic pressure at least ten times for all cases in the data set. Therefore, the magnetic pressure contribution is not sufficient to produce obtained differences in the MP size. We have to note also that there is no correlation ($c$~0.2) between $B_y$ value and the dynamic pressure of SW.

Therefore the influence of $B_y$ on the MP size can not be easy explained by increasing of the IMF magnetic pressure. IMF $B_y$ influence on the MP has both the dynamic and magnetic nature. We believe that the magnetic nature of $B_y$ influence may be explained by reconnection processes in the vicinity of the magnetic cusp region. In this case one can expect the dependence of longitudinal location of the cusp on $B_y$ value. ANN model does not able to represent this effect because we introduce the mirror symmetry of the modeling MP shape relatively equatorial plane.

*Modified pressure balance equation at the subsolar point*
Fig. 5 shows the poor dependence of the subsolar point location on $B_z$ for positive and strong negative $B_z$ values. This is in good agreement with the model of *Kuznetsov et al.* [1998] and conclusions of *Kuznetsov and Suvorova* [1997, 1998a] about existing of different regimes of the magnetopause formation. Let us to discuss the pressure balance at the subsolar point. The pressure balance at subsolar point (thermal and magnetic pressures are neglected) may be written as $kP = B_{ss}^2/8\pi$, where $B_{ss}$ is magnetospheric magnetic field at subsolar point. Theoretical studies show that coefficient $k$ depends on the parameters of the interplanetary medium: sonic Mach number [*Spreiter, et al.*, 1966] and directions of IMF and SW velocity vectors [*Lees*, 1964].

In the case of dipole magnetic field $B_{ss}\sim R^{-3}$ therefore $P\sim R^{-6}$. Real magnetosphere magnetic field is provided by additional sources: ring current, cross-tail current and field aligned currents. Therefore the magnetic field on the magnetopause differs from the double dipole magnetic field due to influence of these additional sources [*Schield*, 1969]. These currents are controlled by interplanetary medium conditions. So in general we can write the expression for the MP magnetic field in the subsolar point as $B_{ss} = \eta R_{ss}^{-1/2b}$, where $\eta$ and $b$ will be depended on the external conditions. Therefore we can rewrite the pressure balance equation at the subsolar point in modified form:

$$R_{ss} = a \cdot P^{-b} \qquad (2)$$

where $a=(\eta^2/8\pi k)^b$. Traditionally the coefficients $a$ and $b$ are considered as functions of SW and IMF parameters. In the model of *Shue et al.* [1997] coefficient $b$ is a constant $b=1/6.6$. In the model of *Kuznetsov et al.* [1998] the power order $b$ weakly and non monotonically varied with $B_z$: $b=0.175\div0.2$ when $-6<B_z(nT)<0$. When $B_z$ is less than -6 nT or $B_z$ is positive than $b=0.19$. The main dependence of the pressure balance on $B_z$ usually performed in coefficient $a=a(B_z)$. This dependence has a linear form in the model of *Shue et al.* [1997]. The model of *Kuznetsov et al.* [1998] has a very complicated dependence $a(B_z)$: $a$ is a constant under positive $B_z$ and exponentially decreased till 8.51 with negative $B_z$.

Using ANN model we have calculated the dependence $R_{ss}(P)$ for various values of $B_z$ in the range $-20\div20$ nT and for two values of $B_y$ (0 nT and 20 nT). Under any conditions the dependence $R_{ss}(P)$ has a good agreement with power low. The coefficients $a$ and $b$ are calculated by simple power approximation. In Fig. 6 we present the results of the calculations for $a$ (down panel) and $b$ (upper panel) versus $B_z$ at $B_y=0$ nT (circles) and at $B_y=20$ nT (triangles). The model dependencies $a(B_z)$ and $b(B_z)$ can be described by modified logistic function:

$$Y(x) = Y_0 + \frac{\alpha}{1+exp\{\beta(x-\gamma)\}} \qquad (3)$$

where variables $Y_0$ and $\alpha$ may be estimated as asymptotic of the function $Y(x)$ when $x\to-\infty$ and $x\to+\infty$ respectively.

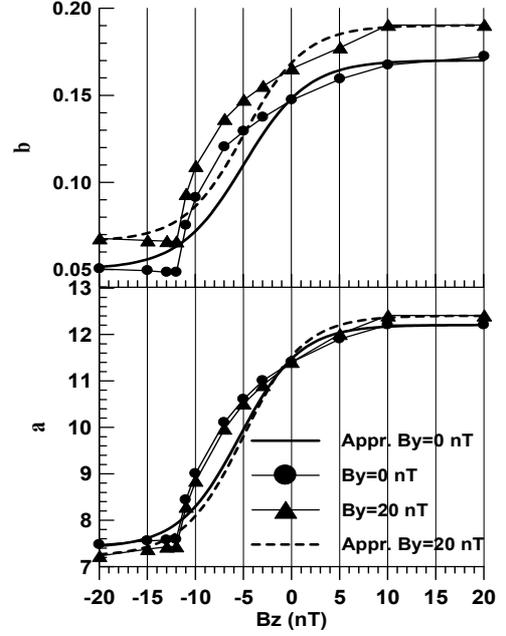

**Figure 6.** Approximation of the coefficient $a(B_z)$ (bottom panel) and power $b(B_z)$ (top panel) of modified pressure balance equation at subsolar point by modified logistic function for $B_y=0$ nT (hard solid lines) and for $B_y=20$ nT (hard dashed lines).

The other coefficients in equation (3) ($\beta$ and $\gamma$) are calculated by simple approximation. The approximations of $a(B_z)$ and of $b(B_z)$ are shown in Fig. 6 for $B_y=0$ nT and for $B_y=20$ nT by hard solid and hard dashed curves respectively. Therefore we obtain following subsolar distance representations in the modified pressure balance equation:

for $B_y=0$nT

$$R_{ss} = \left\{12.2 - \frac{4.7}{1+e^{0.32\cdot(Bz+5.4)}}\right\} \cdot P^{-\left\{0.17 - \frac{0.12}{1+e^{0.3\cdot(Bz+5.1)}}\right\}} \qquad (4a)$$

for $B_y=20$nT

$$R_{ss} = \left\{12.4 - \frac{5.2}{1+e^{0.32\cdot(Bz+5.0)}}\right\} \cdot P^{-\left\{0.19 - \frac{0.12}{1+e^{0.32\cdot(Bz+4.9)}}\right\}} \qquad (4b)$$

We would like to indicate that these dependencies have three regimes: the first one is a regime of positive $B_z$, the second one is an intermediate regime ($-10<B_z(nT)<0$) and the third regime is associated with strong negative $B_z$ ($B_z<-10$ nT). We would like to note that the inflection point $\gamma$ and the scale factor β are similar for $a(B_z)$ and for $b(B_z)$ independently on $B_y$ values: γ≈5 nT, β≈0.32. The power order in the equations (4a) and (4b) is slightly increased with $B_y$.

The results of calculations using expression for $B_y$=0 nT (4a) and for $B_y$=20 nT (4b) are presented in Fig. 7 as solid and dashed lines respectively. Under the positive $B_z$ (the first regime) the equation (4a) is transformed practically into the classic pressure balance equation [*Chapman and Ferraro*, 1931; *Spreiter et al.*, 1966] when the $B_z$ influence is neglected: $R_{ss} \approx 12 \cdot P^{-1/6}$. In the first regime magnetic field at the subsolar point is proportional to the dipole magnetic field: $B_{ss} \sim R_{ss}^{-3}$.

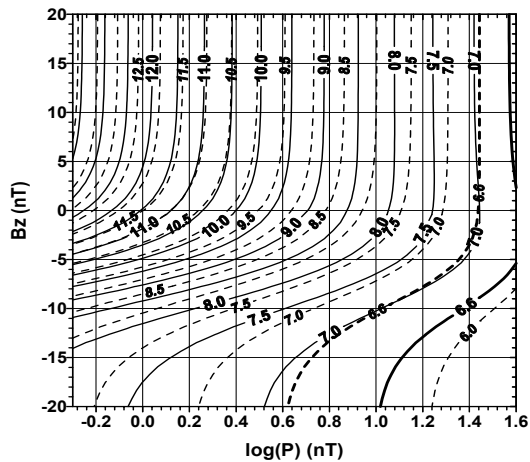

**Figure 7.** The calculations of subsolar point distance (in the Earth's radii) as a bivariate function of P and $B_z$ component using expression (4a) for $B_y$=0 nT (solid lines) and (4b) for $B_y$=20 nT (dashed lines).

In the second regime $R_{ss}$ depends strongly both on the dynamic pressure $P$ and on the $B_z$. The intermediate regime is associated with reconfiguration process of the magnetosphere current system due to dayside reconnection of geomagnetic field and IMF under negative $B_z$. It is important to note that the most probable interplanetary conditions are characterized by -2<$B_z$<1 nT (see Fig. 2a). Therefore the intermediate regime of MP is observed most frequently.

In the third regime the dependence $R_{ss}(B_z)$ has a tendency of disappearing due to probably saturation of reconnection process. This regime is also characterized by very weak dependence on SW dynamic pressure: $R_{ss} = 7.5 \cdot P^{-1/20}$ (for $B_y$=0 nT). Therefore we obtain stronger $B_{ss}(R_{ss})$ dependence: $B_{ss} \sim R_{ss}^{-10}$. Such dependence is probably connected with changing of the global magnetospheric current system configuration that restricts the merging process. Thus the main property of the third regime is week dependence of $R_{ss}$ both on the SW dynamic pressure $P$ and on the IMF $B_z$ component (reconnection saturation).

## Conclusion

3-D model of dayside magnetopause is developed by means of Artificial Neural Networks implemented in NeuroShell 2 package. ANN model is in well agreement with previous MP models. Furthermore ANN model permits to study the influence of many more factors than traditionally used $P$ and $B_z$ and to reconstruct the real three-dimensional geometry of the magnetopause surface. The model may be applied in the extend ranges of solar wind and IMF parameter values: $B_y$=-20÷20 nT; $B_z$=-20÷20 nT; $P$=0.5÷30 nPa. The average relative deviation is better than ≈11% in the wide range of the external parameter values.

The magnetopause shape both in meridian and in equatorial sections is presented. ANN model well described the following main features of MP surface: the cusp region; erosion "dimple" near the equator plane at $B_z$<-10 nT and dawn-dusk asymmetry of MP shape under any SW and IMF conditions.

ANN model quantitatively defines the contribution of $B_y$ IMF influence on the magnetopause dynamics. As expected this contribution is relatively small in comparison with $B_z$ and $P$. The magnetopause asymmetry does not depend on $B_y$.

The modified pressure balance equation at subsolar point is deduced in the form of power low of SW dynamic pressure $P$ with coefficients presented via modified logistic functions of $B_z$. This equation describes three regimes of the magnetopause dynamics: ordinary pressure balance regime ($B_z$>0 nT), intermediate regime with effective reconnection process (0>$B_z$>-10 nT) and regime of the reconnection saturation ($B_z$<-10nT).

**Acknowledgements.** We thank Dr. J.King and others of Goddard Space Flight Center for maintaining the OMNI on line database of solar wind parameters. We thank Prof. I.G. Persiantsev, S.A. Dolenko, Yu.V. Orlov and Ju.S. Shugai for helpful comments and suggestions in practical aspects of ANN usage. We thank referees for very important and useful critics and recommendations.

# Appendix

3D ARTIFICIAL NEURAL NETWORK MODEL OF THE DAYSIDE MAGNETOPAUSE
by A. Dmitriev and A. Suvorova

Skobeltsyn Institute of Nuclear Physics Moscow State University, Moscow, 119899, Russia
E-mails: alexei_dmitriev@yahoo.com 'A. Dmitriev'
     suvorova_alla@yahoo.com 'A. Suvorova'
  Tel: +7 (095) 939-4290
  Fax: +7 (095) 939-5034

Optimal dynamic range of the model (relative error <10%):
-90<vLon<90 (degrees)
-80<vLat<80 (degrees)
-20<By<20 (nT)
-20<Bz<20 (nT)
0.5<Pd<40 (nPa) - solar wind dynamic pressure (see program)

The directory contains following files:
mp_in.dat  - initial data set of the model;
mp_out.dat - output file of the model;
mmp3d.for  - FORTRAN Subroutine with program code of the model;
mp.for     - program for applying of the model code for initial data set to obtain the output file

mp_in.dat format description:
Long    - Longitude GSM (in degrees)
Lat     - Latitude GSM (in degrees)
By(GSM) - Y-IMF component GSM (nT)
Bz(GSM) - Z-IMF component GSM (nT)
Dens    - density of solar wind plasma (1/cm^3)
Vsw     - velocity of solar wind (km/s)
Re      - radial distance of the magnetopause (in the Earth's radii)

mp_out.dat format description:
NN      - Number of the example
Long    - Longitude GSM (in degrees)
Lat     - Latitude GSM (in degrees)
By(GSM) - Y-IMF component GSM (nT)
Bz(GSM) - Z-IMF component GSM (nT)
Dens    - density of solar wind plasma (1/cm^3)
Vsw     - velocity of solar wind (km/s)
Pd      - solar wind dynamic pressure (nPa)
Re      - radial distance of the magnetopause (in the Earth's radii)
Rmod    - ANN modeled radial distance of the magnetopause (in the Earth's radii)
dr      - relative error of the modeled magnetopause distance (dr=(Rmod-Re)/Re)

mp.for
The program calculates also the root mean square deviation (RMSD) for ANN modeled distances:
$RMSD = SQRT(Sum((Re-Rmod)^2))/N)$, where N is total number of the examples.

```
C*******************************************************************
C***************  Artificial Neural Network 3D model  ***************
C***************  of the dayside Earth's magnetopause ***************
C*******************************************************************
       SUBROUTINE MMP3D
     *(
     * vLon,      !in R*4 - Longitude GSM (degrees)
     * vLat,      !in R*4 - Latitude GSM (degrees)
     * By,        !in R*4 - Y-IMF component GSM (nT)
     * Bz,        !in R*4 - Z-IMF component GSM (nT)
     * dens,   !in R*4 - density of solar wind plasma (1/cm^3)
     * V,!in R*4 - velocity of solar wind (km/s)
     * R  !out R*4- radial distance of the magnetopause (in the Earth's radii)
     *)

C Remarks:
C 1. Program returns R=0 in the case of error in the input data
C    (some of the input parameters is out of the dynamic range of the model)
C
C 2. Optimal dynamic range of the model (relative error <10%):
C    -90<vLon<90 (degrees)
C    -80<vLat<80 (degrees)
C    -20<By<20 (nT)
C    -20<Bz<20 (nT)
C    0.5<Pd<40 (nPa) - solar win dynamic pressure (see program)
C
C 3. The model can operate in expanded dynamic range with unexpected accuracy
C    (please change manually):
C    for By: -70 - 40 (nT)
C    for Bz: -40 - 40 (nT)
C    for dynamic pressure (see program): 0.01 - 60 nPa
C
C 4. Please send any comments and recommendations to Dr. A. Dmitriev
C    alexei_dmitriev@yahoo.com 'A. Dmitriev' (SINP MSU, Moscow, 119899, Russia)

       dimension p(24)
       pi=acos(-1.)

       R=0.                                             ! Error value
       Pd=1.672E-06*dens*V*V                            ! Dynamic pressure (nPa)

C Checking on dynamic range
       if(vLon.lt.-90.or.vLon.gt.90) then               ! -90<vLon<90 degrees
         print *,' ERROR! Longitude=',vLon
         pause ' Longitude is out of the dynamic range -90-90 degs'
         return
       endif
       if(vLat.lt.-80.or.vLat.gt.90) then               ! -80<vLat<90 degrees
         print *,' ERROR! Latitude=',vLat
         pause ' Latitude is out of the dynamic range -80-80 degs'
         return
       endif
       if(By.lt.-20.or.By.gt.20) then                   ! -20<By<20 nT
         print *,' ERROR! IMF By=',By
         pause ' By (GSM) is out of the dynamic range -20-20 nT'
         return
       endif
       if(Bz.lt.-20.or.By.gt.20) then                   ! -20<Bz<20 nT
         print *,' ERROR! IMF Bz=',Bz
         pause ' Bz (GSM) is out of the dynamic range -20-20 nT'
         return
       endif
       if(Pd.lt.0.5.or.Pd.gt.40) then                   ! 0.5<Pd<40 nPa
         print *,' ERROR! Pressure=',Pd
         pause ' Pressure is out of the dynamic range 0.5-40 nPa'
         return
       endif
```

```
C Normalisation
      p(1)=vLon/180.                              ! Lon
      p(2)=p(1)**2                                ! Lon^2
      p(3)=p(1)**3                                ! Lon^3
      p(4)=sin(0.5*vLon*pi/180.)      ! sin(1/2Lon)
      p(5)=cos(0.5*vLon*pi/180.)      ! cos(1/2Lon)
      p(6)=sin(2.*vLon*pi/180.)       ! sin(2Lon)
      p(7)=cos(2.*vLon*pi/180.)              ! cos(2Lon)
      p(8)=sin(3.*vLon*pi/180.)       ! sin(3Lon)
      p(9)=cos(3.*vLon*pi/180.)       ! cos(3Lon)
      p(10)=vLat/90.                              ! Lat
      p(11)=p(10)**2                              ! Lat^2
      p(12)=p(10)**3                              ! Lat^3
      p(13)=sin(4.*vLat*pi/180.)      ! sin(4Lat)
      p(14)=cos(4.*vLat*pi/180.)      ! cos(4Lat)
      p(15)=sin(5.*vLat*pi/180.)      ! sin(5Lat)
      p(16)=cos(5.*vLat*pi/180.)      ! cos(5Lat)
      p(17)=By/20.                                ! By
      p(18)=p(17)**2                              ! By^2
      p(19)=p(17)**3                              ! By^3
      p(20)=Bz/20.                                ! Bz
      if(Bz.lt.-12) then              ! ln(Bz)
        p(21)=-1.
      else
        p(21)=(alog(13.+Bz)-1.5)/1.5
      endif
      p(22)=alog(Pd)/4.               ! ln(Pd)
      p(23)=exp(p(22))-1.4                        ! Pd
      p(24)=0.                        ! ln(R)

C ANN Normalisation
      X1=2.*(p(1)+.5)-1.
      X2=2.*p(2)/.25-1.
      X3=2.*(p(3)+.125)/.25-1.
      X4=2.*(p(4)+.70711)/1.41422-1.
      X5=2.*(p(5)-.70711)/.29289-1.
      X6=2.*(p(6)+1.)/2.-1.
      X7=2.*(p(7)+1.)/2.-1.
      X8=2.*(p(8)+1.)/2.-1.
      X9=2.*(p(9)+.98769)/1.98769-1.
      X10=2.*(p(10)+.8889)/1.7778-1.
      X11=2.*p(11)/.79014-1.
      X12=2.*(p(12)+.70236)/1.40472-1.
      X13=2.*(p(13)+.9945)/1.989-1.
      X14=2.*(p(14)+.97819)/1.97819-1.
      X15=2.*(p(15)+.98484)/1.96968-1.
      X16=2.*(p(16)+.93981)/1.93981-1.
      X17=2.*(p(17)+.5)/1.5-1.
      X18=2.*p(18)-1.
      X19=2.*(p(19)+.125)/1.125-1.
      X20=2.*(p(20)+.6)/.9543-1.
      X21=2.*(p(21)+1)/2.-1.
      X22=2.*(p(22)+.125)/.9-1.
      X23=2.*(p(23)+.5175)/1.28809-1.

C Calculation Expression
      sum=-3.139507548628609E-003*X10
      sum=sum-0.2547149526398687*X22
      sum=sum-0.4011627537775905*X2
      sum=sum-0.1973017399439671
      sum=sum+3.296012272125294E-003*X18
      sum=sum-3.056257543867325E-002*X3
      sum=sum+5.138537479870376E-004*X9
      sum=sum-4.648953277724813E-002*X23
      sum=sum-1.891764181319954E-003*X14
      sum=sum+2.43759400112602E-002*X6
      sum=sum-6.960854888136313E-003*X16
```

```
sum=sum+0.1047466313549833*X21
sum=sum-0.2882329794547871*X4
sum=sum+0.2037198844933279*X11
sum=sum-1.25817540410462E-002*X19
sum=sum-1.026846522094714E-002*X20
sum=sum-6.166094204329592E-002*X17
sum=sum-0.4265110190468048*X5
sum=sum-8.233199682348016E-002*X5**2
sum=sum-3.110084678103177E-002*X23**2
sum=sum+0.1294582100451645*X5*X23
sum=sum+6.872712883781075E-003*X21*X23
sum=sum-0.1006212556723966*X5*X21*X23
sum=sum+2.25106943541424E-002*X11**2
sum=sum-1.792385161461331E-002*X22**2
sum=sum-0.1227295260008636*X11*X21
sum=sum+0.1643039329788456*X11*X22
sum=sum-9.117921937006793E-002*X21*X22
sum=sum+5.227493586802182E-002*X11*X21*X22
sum=sum+0.4713960554575409*X4**2
sum=sum-5.702587135835827*X4*X5
sum=sum+1.944579333121296E-002*X4*X21
sum=sum-0.223035595627996*X4*X23
sum=sum-1.100804820945859*X4*X5**2
sum=sum+2.097413211764071E-002*X4*X23**2
sum=sum+1.730897187326532*X4*X5*X23
sum=sum-9.030787574270464E-002*X4*X21*X23
sum=sum-1.345337992606676*X4*X5*X21*X23
sum=sum-0.163655195385247*X4*X11
sum=sum+0.1978065906504018*X4*X22
sum=sum-6.535665347289167E-003*X4*X11**2
sum=sum+2.704860704783475E-002*X4*X22**2
sum=sum+9.698863889129741E-002*X4*X11*X21
sum=sum-0.122150921471858*X4*X11*X22
sum=sum+7.205558980005908E-002*X4*X21*X22
sum=sum-4.131096275833116E-002*X4*X11*X21*X22
sum=sum+6.703932062802843E-002*X2**2
sum=sum+4.684613665381818E-002*X2*X23
sum=sum-7.260871222695957E-002*X2*X21*X23
sum=sum-5.508678452432155*X2*X4
sum=sum+1.22101726528913*X2**2*X4
sum=sum+1.635476216868507*X2*X4*X23
sum=sum-1.300085475169895*X2*X4*X21*X23
sum=sum-3.110899595562811E-002*X17**2
sum=sum-1.023042598651039E-002*X2*X17*X23
sum=sum-7.076423531975327E-002*X2*X5*X23
sum=sum-2.656418130606105E-003*X2*X23**2
sum=sum-1.366004754245885E-002*X2*X5**2*X23
sum=sum+2.508556672359299E-004*X2*X23**3
sum=sum+2.147895559693557E-002*X2*X5*X23**2
sum=sum-1.088027591125269E-003*X2*X21*X23**2
sum=sum-1.669449532742118E-002*X2*X5*X21*X23**2
sum=sum+3.435908196649176E-002*X2*X11*X23
sum=sum-4.152909930949929E-002*X2*X22*X23
sum=sum+1.372149908497884E-003*X2*X11**2*X23
sum=sum-5.678801118707159E-003*X2*X22**2*X23
sum=sum-2.036257135399783E-002*X2*X11*X21*X23
sum=sum+2.564534241185707E-002*X2*X11*X22*X23
sum=sum-1.512792741016342E-002*X2*X21*X22*X23
sum=sum+8.67315426306433E-003*X2*X11*X21*X22*X23
sum=sum+7.8211130031236123E-002*X2*X4**2*X23
sum=sum-0.9461401933144435*X2*X4*X5*X23
sum=sum-3.700477284027411E-002*X2*X4*X23**2
sum=sum-0.182639153296959*X2*X4*X5**2*X23
sum=sum+3.479906390501594E-003*X2*X4*X23**3
sum=sum+0.2871804253780189*X2*X4*X5*X23**2
sum=sum-1.4983359127663E-002*X2*X4*X21*X23**2
sum=sum-0.2232106793071529*X2*X4*X5*X21*X23**2
sum=sum-2.715272112646388E-002*X2*X4*X11*X23
```

```
            sum=sum+3.281892261509672E-002*X2*X4*X22*X23
            sum=sum-1.084359699874423E-003*X2*X4*X11**2*X23
            sum=sum+4.487748050407266E-003*X2*X4*X22**2*X23
            sum=sum+1.609179261343579E-002*X2*X4*X11*X21*X23
            sum=sum-2.026657264536157E-002*X2*X4*X11*X22*X23
            sum=sum+1.195504567293592E-002*X2*X4*X21*X22*X23
            sum=sum-6.854075415095674E-003*X2*X4*X11*X21*X22*X23
            sum=sum-6.843874415479644E-002*X2**2*X23
            sum=sum+1.516967979694239E-002*X2**3*X23
            sum=sum+2.090410004010033E-002*X2**2*X23**2
            sum=sum-1.619209111272105E-002*X2**2*X21*X23**2
            sum=sum-0.9139679888692046*X2**2*X4*X23
            sum=sum+0.2025841050566612*X2**3*X4*X23
            sum=sum+0.2791646531321127*X2**2*X4*X23**2
            sum=sum-0.2162379385046533*X2**2*X4*X21*X23**2
            sum=sum-5.161424235381224E-003*X2*X17**2*X23
            sum=sum-2.104608777389302E-002*X20**2
            sum=sum+8.756588363311123E-002*X6*X20
            sum=sum-4.63870207825729E-002*X11*X19
            sum=sum-3.310240865737497E-002*X14**2
            sum=sum+4.64816307915491E-002*X4*X14
            sum=sum+9.244583038920572E-003*X16**2
            sum=sum+2.419584183188094E-002*X21**2
            sum=sum+4.463856244911715E-002*X16*X21
            sum=sum-0.1231258390692942*X3*X9
            sum=sum-2.447847515006272E-002*X6*X11
            sum=sum-3.659676220220129E-002*X18*X22
            sum=sum+3.300114439659334E-002*X9*X14
            sum=sum-2.832437304761224E-002*X9*X23
            sum=sum+3.083814309063122E-002*X3*X20
            sum=sum+1.928841007283686E-002*X8*X23
            sum=sum+3.732925804288816E-002*X3*X22
            sum=sum+5.882703950494068E-003*X18**2
            sum=sum+1.356769996071952E-002*X18*X20
            sum=sum+1.175164044624989E-002*X8*X22
            sum=sum+4.99605904531362E-003*X8**2
            sum=sum-1.147759744909048E-002*X2*X22
            sum=sum+6.770361103064851E-003*X20*X22
            sum=sum-5.639051339293889E-003*X10*X22
            sum=sum-6.566020775898155E-003*X10**2
            p(24)=1.18359*(sum+1.)/2.-.62483

C Restoring from ANN
            R=exp(p(24)+2.4)

            return
            end
```